\documentclass[prx,nofootinbib,aps,twocolumn,showpacs,showkeys,groupaddress,preprintnumbers,floatfix]{revtex4-1}

\usepackage{dynlearn}

\newcommand{\kB}{k_\textrm{B}}

\begin{document}

\def\ourTitle{%
Extreme Quantum Advantage for Rare-Event Sampling}

\def\ourAbstract{%
We introduce a quantum algorithm for efficient biased sampling of the rare
events generated by classical memoryful stochastic processes. We show that this
quantum algorithm gives an extreme advantage over known classical biased
sampling algorithms in terms of the memory resources required. The quantum
memory advantage ranges from polynomial to exponential and when sampling the
rare equilibrium configurations of spin systems the quantum advantage diverges.
}

\def\ourKeywords{%
quantum algorithm, large deviation theory, biased sampling, quantum memory, quantum advantage, stochastic process, hidden Markov model
}

\hypersetup{
  pdfauthor={James P. Crutchfield},
  pdftitle={\ourTitle},
  pdfsubject={\ourAbstract},
  pdfkeywords={\ourKeywords},
  pdfproducer={},
  pdfcreator={}
}

\title{\ourTitle}

\author{Cina Aghamohammadi}
\email{caghamohammadi@ucdavis.edu}

\author{Samuel P. Loomis}
\email{sloomis@ucdavis.edu}

\author{John R. Mahoney}
\email{jrmahoney@ucdavis.edu}

\author{James P. Crutchfield}
\email{chaos@ucdavis.edu}

\affiliation{Complexity Sciences Center and Physics Department,
University of California at Davis, One Shields Avenue, Davis, CA 95616}

\date{\today}

\begin{abstract}
\ourAbstract
\end{abstract}


\preprint{Santa Fe Institute Working Paper 2017-07-XXX}
\preprint{arxiv.org:1707.XXXXX [physics.gen-ph]}

\maketitle

\bibliographystyle{unsrt}

\setstretch{1.1}

\section{Introduction}

From earthquakes to financial market crashes, rare events are associated with
catastrophe---from decimated social infrastructure and the substantial loss of
life to global economic collapse. Though rare, their impact cannot be ignored.
Prediction and modeling such rare events is essential to mitigating their
effects. However, this is particularly challenging, often requiring huge
datasets and massive computational resources, precisely because the events of
interest are rare.

Ameliorating much of the challenge, \emph{biased} or \emph{extended sampling}
\cite{Leac01,Fren07a} is an effective and now widely-used method for efficient
generation and analysis of rare events. The underlying idea is simple to state:
transform a given distribution to a new one where previously rare events are
now typical. This concept was originally proposed in 1961 by Miller to probe
the rare events generated by discrete-time, discrete-value Markov stochastic
processes \cite{Mill61a}. It has since been extended to address non-Markovian
processes \cite{Youn93a}. The approach was also eventually adapted to
continuous-time first-order Markov processes \cite{Leco05,Leco07,Chet13}.
Today, the statistical analysis of rare events is a highly developed toolkit
with broad applications in sciences and engineering \cite{Vara84a}. Given this,
it is perhaps not surprising that the idea and its related methods appear under
different appellations, depending on the research arena. For example, large
deviation theory refers to the \emph{s-ensemble method} \cite{Garr09a,Hedg09a},
the \emph{exponential tilting algorithm} \cite{Vanc81a,Csis84a}, or as
generating \emph{twisted distributions}.

In 1997, building on biased sampling, Torrie and Valleau introduced
\emph{umbrella sampling} into Monte Carlo simulation of systems whose energy
landscapes have high energy barriers and so suffer particularly from poor
sampling \cite{Torr77a}. Since then, stimulated by computational problems
arising in statistical mechanics, the approach was generalized to
\emph{Ferrenberg-Swendsen reweighting}, later still to \emph{weighted histogram
analysis} \cite{Kuma92a}, and more recently to \emph{Wang-Landau sampling}
\cite{Wang01a}. 

When generating samples for a given stochastic process one can employ
alternative types of algorithm. There are two main types---Monte Carlo or
finite-state machine algorithms. Here, we consider finite-state machine
algorithms based on Markov chains (MC) \cite{Levi9a,Norr98} and hidden Markov
models (HMM) \cite{Uppe97a,Rabi86a,Rabi89a}. For example, if the process is
Markovian one uses MC generators and, in more general cases, one uses HMM
generators.

When evaluating alternative approaches the key questions that arise concern
algorithm speed and memory efficiency. For example, it turns out there are HMMs
that are always equally or more memory efficient than MCs. There are many
finite-state HMMs for which the analogous MC is infinite-state \cite{Crut01a}.
And so, when comparing all HMMs that generate the same process, one is often
interested in those that are most memory efficient. For a generic stochastic
process, the most memory efficient classical HMM known currently is the
\emph{\eM} of computational mechanics \cite{Crut12a}. The memory it requires is
called the process' \emph{statistical complexity} $\Cmu$ \cite{Crut88a}.

Today, we have come to appreciate that several important mathematical problems
can be solved more efficiently using a quantum computer. Examples include
quantum algorithms for integer factorization \cite{Shor99aa}, search
\cite{Grov96aa}, eigen-decomposition \cite{Abra99aa}, and solving linear systems
\cite{Harr09aa}. Not long ago and for the first time, Ref. \cite{Gu12a}
provided a quantum algorithm that can perform stochastic process
sample-generation using less memory than the best-known classical algorithms.
Recently, using a stochastic process' higher-order correlations, a new quantum
algorithm---the \emph{q-machine}---substantially improved this efficiency and
extended its applicability \cite{Maho16}. More detailed analysis and a
derivation of the closed-form quantum advantage of the q-machine is given in a
sequel \cite{Riec16a}. Notably, the quantum advantage has been verified
experimentally for a simple case \cite{Pals15}.

The following brings together techniques from large deviation theory, classical
algorithms for stochastic process generation, computational complexity theory,
and the newly introduced quantum algorithm for stochastic process generation to
propose a new, memory efficient quantum algorithm for the biased sampling
problem. We show that there can be an extreme advantage in the quantum
algorithm's required memory compared to the best known classical algorithm. Two
examples are analyzed here. The first is the simple, but now well-studied
\emph{perturbed coin process}. The second is a more physical example---a
stochastic process that arises from the Ising next-nearest-neighbor spin system
in contact with thermal reservoir.

\section{Classical Algorithm}

The object for which we wish to generate samples is a discrete-time,
discrete-value \emph{stochastic process} \cite{Trav14a,Uppe97a}: a
probability space $\mathcal{P}=\big\{\MeasAlphabet^\infty, \Sigma,
\mathbb{P}(\cdot) \big\}$, where $\mathbb{P}(\cdot)$ is a probability measure
over the bi-infinite chain $\ldots \MeasSymbol_{-2} \MeasSymbol_{-1}
\MeasSymbol_0 \MeasSymbol_1 \MeasSymbol_2 \ldots$, each random variable
$\MeasSymbol_i$ takes values in a finite, discrete alphabet $\MeasAlphabet$,
and $\Sigma$ is the $\sigma$-algebra generated by the cylinder sets in
$\MeasAlphabet^\infty$. For simplicity we consider only ergodic stationary
processes: that is, $\mathbb{P}(\cdot)$ is invariant under time
translation---$\mathbb{P}(\MeasSymbol_{i_1}\MeasSymbol_{i_2} \cdots
\MeasSymbol_{i_m}) = \mathbb{P}(\MeasSymbol_{i_1+n}\MeasSymbol_{i_2+n} \cdots
\MeasSymbol_{i_m+n})$ for all $n$, $m$---and over successive realizations.

\begin{figure}
\includegraphics[width=0.8\columnwidth]{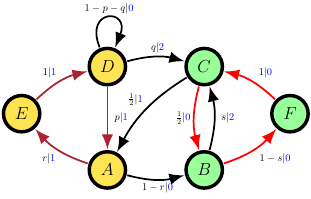}
\caption{Hidden Markov model generator of a stochastic process with
	infinite-range statistical dependencies that requires an HMM with only six
	states. To generate the same process via a Markov chain requires one with
	an infinite number of states and so infinite memory.
  }
\label{fig:HmmExample}
\end{figure}

\emph{Sampling} or \emph{generating} a given stochastic process refers to
producing a finite realization that comes from the process' probability
distribution. Generally, generating a process via its probability measure
$\mathbb{P}(\cdot)$ is impossible due to the vast number of allowed
realizations and, as a result, this prosaic approach requires an unbounded
amount of memory. Fortunately, there are more compact ways than specifying
in-full the probability measure on the sequence sigma algebra. This recalls the
earlier remark that HMMs can be arbitrarily more compact than alternative
algorithms for the task of generation.

An HMM is specified by a tuple $\big\{ \CausalStateSet, \MeasAlphabet, \{
T^{(\meassymbol)},\meassymbol \in \MeasAlphabet \} \big\}$. In this,
$\CausalStateSet$ is a finite set of states, $\MeasAlphabet$ is a finite
alphabet, and $\{T^{(\meassymbol)}, \meassymbol \in \MeasAlphabet\}$ is a set
of $|\CausalStateSet| \times |\CausalStateSet|$ substochastic symbol-labeled
transition matrices whose sum $T= \sum_{\meassymbol \in \MeasAlphabet}
T^{(\meassymbol)}$ is a stochastic matrix.

As an example, consider the HMM state-transition diagram shown in
Fig.~\ref{fig:HmmExample}, where $\CausalStateSet= \{ A,B,C,D,E,F\}$,
$\MeasAlphabet = \{0,1,2\}$, and we have three $6 \times 6$ substochastic
matrices $T^{(0)}$, $T^{(1)}$, and $T^{(2)}$. Each edge is labeled
$p|{\color{blue}x}$ denoting the transition probability $p$ and a symbol
${\color{blue}x} \in \MeasAlphabet$ which is emitted during the transition. In
this HMM, of the two edges exiting state $C$, one enters state $B$ and the
other enters state $A$. The edges from $C$ to $A$ and $C$ to $B$ are labeled by
$\frac{1}{2}|{\color{blue}1}$ and $\frac{1}{2}|{\color{blue}0}$. This simply
means that if the HMM is in the state $C$, then with probability $\frac{1}{2}$
it goes to the state $A$ and emits the symbol $\color{blue}1$ and with
probability $\frac{1}{2}$ it goes to state $B$ and emits symbol
$\color{blue}0$. Following these transition rules in succession generates
realizations in the HMM's process.

How does this generation method compare to generating realizations of the same
process via a finite Markov chain. It turns out that this cannot be
implemented, since generating a symbol can depend on the infinite history. That
is, the process has infinite Markov order. As a result, to generate a
realization using a Markov chain one needs an infinite number of Markovian
states. In other words, implementing the Markov chain algorithm to generate
process samples on a conventional computer requires an infinite amount of
memory.

To appreciate the reason behind the process' infinite Markov order, refer to
Fig.~\ref{fig:HmmExample}'s HMM. There are two length-$3$ state-loops
consisting of the edges colored red (right side of state-transition diagram)
and those colored maroon (left side). Note that if the HMM generates $n$
$\color{blue}1$s in a row, we will not know the HMM's current state, only that
it is either $A$, $D$, or $E$. This state uncertainty (entropy) is bounded away
from $0$. The observation holds for the other loop and its sequences of symbol
$\color{blue}0$ and the consequent ambiguity among states $B$, $C$, and $F$.
Thus, there exist process realizations from which we cannot determine the
future statistics, independent of the number of symbols seen. This means that
the process statistics depend on infinite past sequences---the process has
infinite Markov order. To emphasize, implementing a MC algorithm for this
requires infinite memory. The contrast with the finite HMM method is an
important lesson: HMMs are strictly more powerful generators, as a class of
algorithms, than Markov chain generators.

For any given process $\mathcal{P}$, there are an infinite number of HMMs that
generate it. Therefore, one is compelled to ask, Which algorithm requires the
least memory for implementation? The best known implementation, and provably
the optimal predictor, is known as the \eM \cite{Shal98a, Crut12a}. The states
of the \eM are called \emph{causal states}; we denote this set
$\CausalStateSet$.

The average memory required for $M(\mathcal{P})$ to generate process
$\mathcal{P}$ is given by the process' \emph{statistical complexity}
$\Cmu(\mathcal{P})$ \cite{Crut88a}. To calculate it:
\begin{enumerate}
\item Compute the stationary distribution $\pi$ over causal states. $\pi$
	is the left eigenvector of the state-transition matrix $T$ with eigenvalue
	$1$: $\pi T = \pi$.
\item Calculate the state's Shannon entropy $\H[\CausalStateSet] =
-\sum_{\cs \in \CausalStateSet} \pi(\cs) \log_2 \pi(\cs)$.
\end{enumerate}
Thus, $\Cmu = \H[\CausalStateSet]$ measures the (ensemble average) memory required to generate the process.

Another important, companion measure is $\hmu$, the process' \emph{metric
entropy} (or Shannon entropy rate) \cite{Han06}:
\begin{align*}
\hmu(\mathcal{P}) = - \lim_{n \to \infty}
  \frac{1}{n} \sum_{w \in \MeasAlphabet^n }
  \mathbb{P}(w) \log_2 \mathbb{P}(w)
  ~.
\end{align*}
Although sometimes confused, it is important to emphasize that $\hmu$ describes
randomness in the realizations, while $\Cmu$ describes the required memory for
process generation.

\section{Quantum memory advantage}\label{Sec:QuanAdv}

Recently, it was shown that a quantum algorithm for process generation can use
less memory than the best known classical algorithm (\eM) \cite{Gu12a}. We
refer to the ratio of required classical memory $\Cmu$ to quantum memory as the
\emph{quantum advantage}. Taking into account a process' higher-order
correlations, a new quantum algorithm---the \emph{q-machine}---was introduced
that substantially improves the original quantum algorithm and is, to date, the
most memory-efficient quantum algorithm known for process generation
\cite{Maho16}. Closed-form expressions for the quantum advantage are given in
\cite{Riec16a}.

Importantly, the quantum advantage was recently verified experimentally for the
simple \emph{perturbed coins process} \cite{Pals15}. It has been found that the
\emph{q-machine} sometimes confers an extreme quantum-memory advantage. For
example, for generation of ground-state configurations (in a Dyson-type spin
model with $N$-nearest-neighbor interactions at temperature $T$), the quantum
advantage scales as $N T^2 / \log_2{T}$ \cite{Agha16b, Garn16a}. 

One consequence of this quantum advantage arises in model selection
\cite{Agha16a}. Statistical inference of models for stochastic systems often
involves controlling for model size or memory. The following applies this
quantum advantage to find gains in the setting of biased sampling of a process'
rare events. In particular, we will develop tools to determine how the memory
requirements of classical and quantum algorithms vary over rare-event classes.

\section{Quantum algorithm}
\label{Sec:quantumconstruction}

\newcommand{\tr}{\mathrm{tr}}

We define the \emph{quantum machine} of a stochastic process $\mathcal{P}$, by
$QM(\mathcal{P}) = \{ \mathcal{H}, \MeasAlphabet, \{
K_{\meassymbol},\meassymbol \in \MeasAlphabet \} \}$, where $\mathcal{H}$
denotes the Hilbert space in which quantum states reside, $\MeasAlphabet$ is
the same alphabet as the given process', and $\{ K_{\meassymbol},\meassymbol
\in \MeasAlphabet \}$ is a set of Kraus operators we use to specify the
measurement protocol for states \cite{Pres98a}.\footnote{We adopt a particular
form for the Kraus operators. In general, they are not unique.} Assume we have
the state (or density matrix) $\rho_0 \in \mathcal{B}(H)$ in hand. We perform a
measurement and, as a result, we measure $X$. The probability of yielding
symbol $x_0 \in X$ is:
\begin{align*}
\mathbb{P}(X=x_0| \rho_0) = \tr \left( K_{x_0} \rho_0 K_{x_0}^\dagger \right)
  ~.
\end{align*}
After measurement with outcome $X=s_0$, the new quantum state is:
\begin{align*}
\rho_1 = \frac{K_{x_0} \rho_0 K_{x_0}^\dagger}{\tr(K_{x_0} \rho_0 K_{x_0}^\dagger)}
  ~.
\end{align*}
Repeating these measurements generates a stochastic process. The process
potentially could be nonergodic, depending on the initial state $\rho_0$.
Starting the machine in the stationary state defined by:
\begin{align*}
\rho_s = \sum \limits_{x \in \MeasAlphabet} K_x \rho_s K_x^\dagger
  ~, 
\end{align*}
and doing a measurements over and over again leads to generating a stationary
stochastic process over $\meassymbol \in \MeasAlphabet$. For any given process,
$\rho_s$ can be calculated by the method introduced in Ref.~\cite{Riec16a}.

Our immediate goal is to design a quantum generator of a given classical
process. (Section~\ref{Sec:BiasedSampling} will then take the given process to
represent a rare-event class of some other process.) For now, we start with the
process' \eM. The construction consists of three steps, as follows.

{\bf First:} Map every causal state $\causalstate_i \in \CausalStateSet$ to a
pure quantum state $\ket{\eta_i}$. Each signal state $\ket{\eta_i}$ encodes the
set of length-$R$ sequences that may follow $\causalstate_i$, as well as each
corresponding conditional probability: 
\begin{align*}
\ket{\eta_i} \equiv
  \sum \limits_{w \in \MeasAlphabet^R}
  \sum \limits_{\causalstate_j \in \CausalStateSet}
  {\sqrt{\mathbb{P}(w| \causalstate_i)}
  ~ \ket{w}}
  ~,
\end{align*}
where $w$ denotes a length-$R$ sequence, $\mathbb{P}(w| \causalstate_i) =
\mathbb{P}(X_{0} \cdots X_{R-1} = w | \CausalState_0 = \causalstate_i)$, and
$R$ is the process' the Markov order. The resulting Hilbert space is
$\mathcal{H}_w $ with size $|\mathcal{A}|^R$, the number of length-$R$
sequences, with basis elements $\ket{w} = \ket{x_0} \otimes \cdots \otimes \ket{x_{R-1}}$.

{\bf Second}: Form a matrix $\Xi$ by assembling the signal states:
\begin{align*}
\Xi = 
\begin{bmatrix}
\ket{\eta_0} && \ket{\eta_1} &&\cdots && \ket{\eta_{|\CausalStateSet|-1}}
\end{bmatrix}~.
\end{align*}
From here on out, we assume all the $\ket{\eta_i}$s are linearly independent.
(This holds for general processes except for some special cases,
which we discuss elsewhere.)
Define $|\CausalStateSet|$ new bra states $\ket{\widetilde{\eta_i}}$:
\begin{align*}
\begin{bmatrix}
\bra{\widetilde{\eta_0}} \\ \bra{\widetilde{\eta_1}} \\ \cdots \\ \bra{\widetilde{\eta_{|\CausalStateSet|-1}}}
\end{bmatrix} = \Xi^{-1}~.
\end{align*}
That is, we design the new bra states such that we obtain the identity:
\begin{align*}
\begin{bmatrix}
\bra{\widetilde{\eta_0}} \\ \bra{\widetilde{\eta_1}} \\ \cdots \\ \bra{\widetilde{\eta_{|\CausalStateSet|-1}}}
\end{bmatrix}
\begin{bmatrix}
\ket{\eta_0} && \ket{\eta_1} &&\cdots && \ket{\eta_{|\CausalStateSet|-1}}
\end{bmatrix}
  = \mathrm{I}
  ~.
\end{align*}

{\bf Third}: Define $|\mathcal{A}|$ Kraus operators $K_i$s via:
\begin{align*}
K_x = \sum_{i,j} \sqrt{T_{ij}^x} \ket{\eta_j}\bra{\widetilde{\eta_i}}
  ~.
\end{align*}

Using the quantum generator $QM(\mathcal{P})$, the required average memory for
generating process  $\mathcal{P}$ is $C_q(\mathcal{P}) = S(\rho_s)$, where
$S(\rho) = -\tr (\rho \log \rho )$ denotes the \emph{von Neumann entropy}
\cite{Pres98a}. References \cite{Maho16,Agha16b} explain why $C_q$ is the
quantum machine's required memory.

\section{Typical Realizations}

At this point, we established classical and quantum representations of
processes and characterized their respective memory requirements. Our purpose
now turns to this set-up to monitor the classical and quantum resources
required to generate probability classes of a process' realizations.

The concept of a stochastic process is quite general. Any physical system that
exhibits stochastic dynamics in time or space may be thought of as
\emph{generating} a stochastic process. In the spatial setting one considers
not time evolution, but rather the spatial ``dynamic''. For example, consider a
one-dimensional noninteracting Ising spin-\textonehalf\ chain with classical
Hamiltonian $ H = -\sum_{i=1}^{n} h \sigma_i$ in contact with a thermal
reservoir at temperature $T$. After thermalizing, a spin configuration at one
instant of time may be thought of as having been generated left-to-right (or
equivalently right-to-left). The probability distribution over these
spatial-translation invariant configurations defines a stationary stochastic
process---a simple Markov random field.

For $n \gg 1$, one can ask for the probability of seeing $k$ up spins. The
Strong Law of Large Numbers \cite{Durr10} guarantees that for large $n$, the
ratio $k/n$ almost surely converges to
$p_{\uparrow} = \tfrac{1}{2} \left(1+ \tanh(h/\kB T) \right)$.
That is:
\begin{align*}
\mathbb{P} \left( \lim_{n \to \infty} \frac{k}{n} = p_{\uparrow} \right) = 1
  ~. 
\end{align*}
Informally, a \emph{typical sequence} is one that has close to $p_{\uparrow} n$
spin ups. However, this does not preclude seeing other kinds of rare long runs,
e.g., all up-spins or all down-spin. It simply means that the latter are
\emph{rare events}.

Now let us formally define the concept of typical realizations and, consequently, rare ones.
Consider a given process $\mathcal{P}$ and let $\MeasAlphabet^n$ denote its set of length-$n$
realizations. Then, for an arbitrary $0 < \epsilon \ll 1$ the process'
\emph{typical set} \cite{Cove06a,Kull68,Yeun08a} is defined:
\begin{align}
A_{\epsilon}^{n} \! \equiv \! \{ w:
  2^{-n (h_\mu + \epsilon)} \leq \mathbb{P}(w)
  \leq 2^{-n (h_\mu - \epsilon)}, w \in \MeasAlphabet^n \}
  ,
\label{eq:TSDEF}
\end{align}
where $\hmu$ is the process' Shannon entropy rate, introduced above.

According to the \emph{Shannon-McMillan-Breiman theorem}
\cite{Shan48a,McMi53a,Brei57}, for a given $\epsilon \ll 1$ and sufficiently large $n^*$:
\begin{align}
\mathbb{P}(w \notin A_{\epsilon}^{n}, w \in \MeasAlphabet^n) \leq \epsilon, \forall n \geq n^*
  ~.
\label{eq:AtypProb}
\end{align}
There are two important lessons here. First, from Eq. (\ref{eq:TSDEF}) we see
that all sequences in the typical set have approximately the same probability.
More precisely, the probability of typical sequences decays at the same
exponential rate. The following adapts this to use decay rates to identify
distinct sets of rare events. Second, coming from Eq. (\ref{eq:AtypProb}), for
large $n$ the probability of sequences falling outside the typical set is close
to zero---these are the sets of rare sequences.

\begin{figure}
\includegraphics[width=0.7\columnwidth]{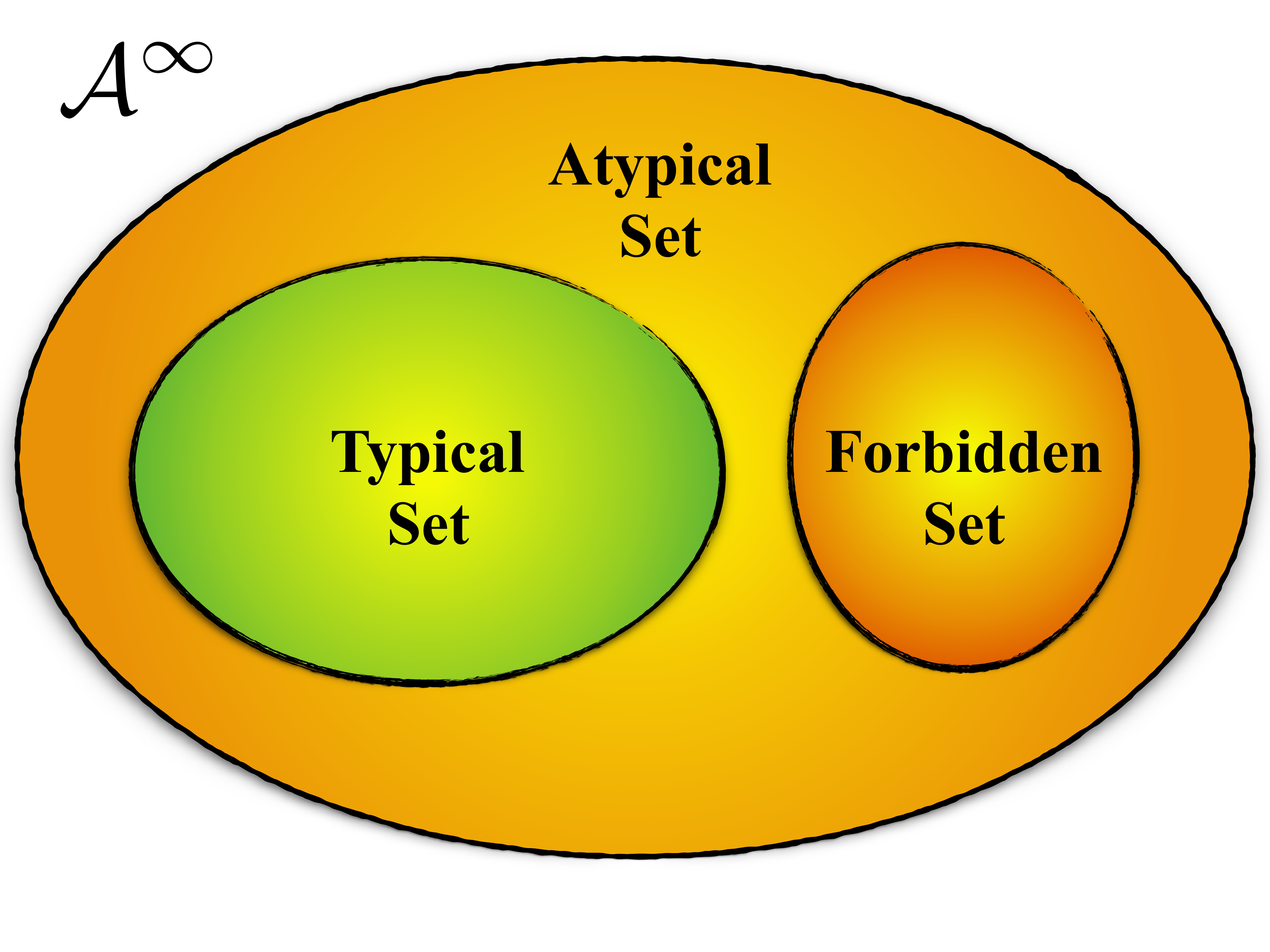}
\caption{For a given process, the space $\MeasAlphabet^\infty$ of all sequences
	is partitioned into those that are forbidden by the process, sequences in
	the typical set, and sequences not forbidden nor typical---the
	\emph{atypical} or rare sequences.
  }
\label{fig:TSNTS}
\end{figure}

Another important consequence of the theorem is that sequences generated by a
stationary ergodic process fall into one of three partitions; see
Fig.~\ref{fig:TSNTS}. The first contains those that are never generated; they
fall in the the \emph{forbidden set}. For example, the HMM in
Fig.~\ref{fig:HmmExample} never generates sequences that have consecutive $2$s.
The second partition consists of those in the typical set---the set with
probability close to one, as in Eq. (\ref{eq:TSDEF}). And, the last contains
sequences in a family of atypical sets---realizations that are rare to
different degrees. We now refine this classification by dividing the atypical
set into identifiable subsets, each with their own characteristic rarity.

\begin{figure}
  \centering
\includegraphics[width=\columnwidth]{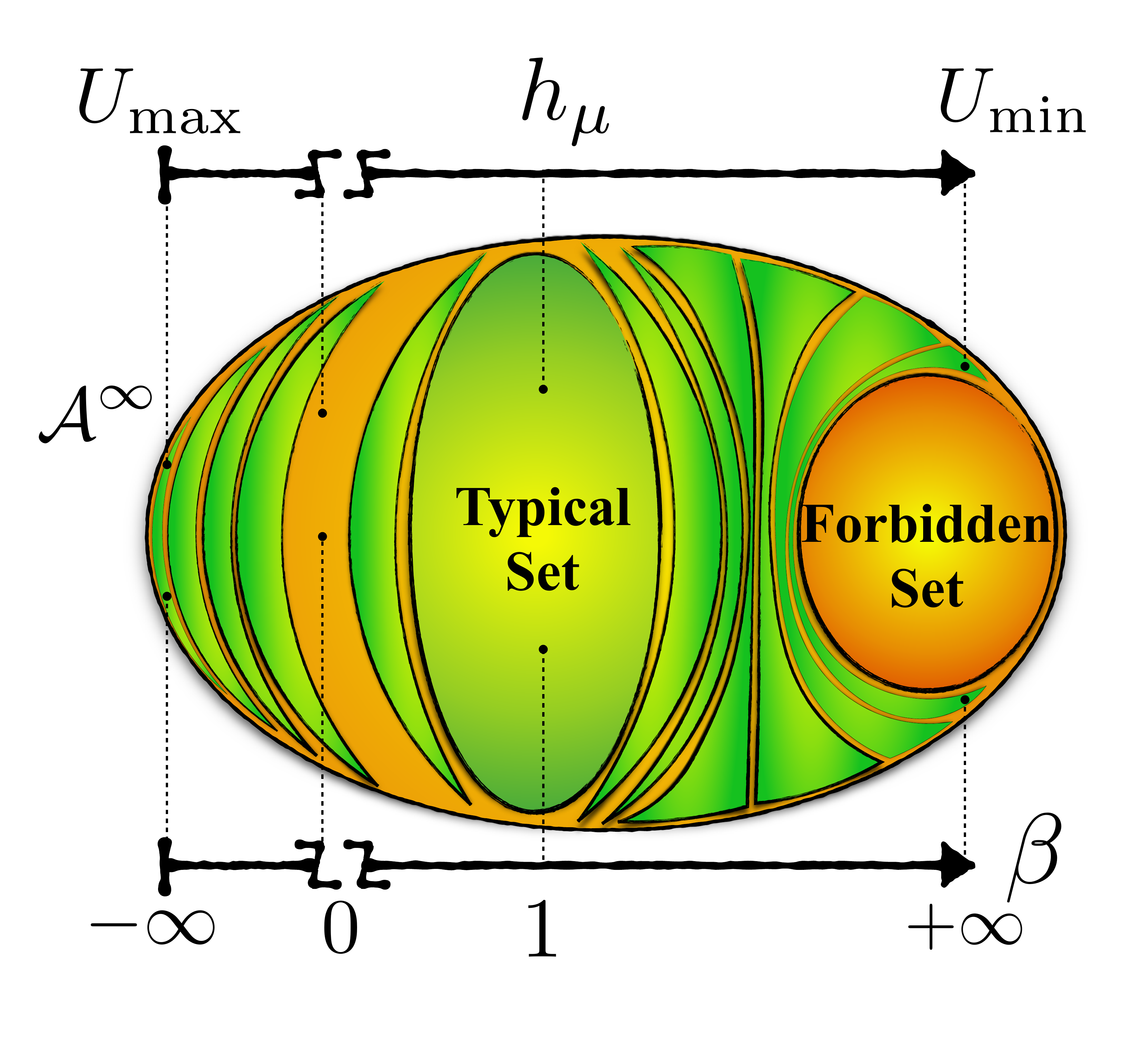}
\caption{The space of all sequences $\MeasAlphabet^\infty$ partitioned into
	$\Lambda_U$s---isoenergy or equal probability-decay-rate bubbles---in which
	all sequences in the same $\Lambda_U$ have the same energy $U$. The typical
	set is one such bubble with energy equal to Shannon entropy rate: $U =
	\hmu$. Another important class is the forbidden set, in which all sequences
	do not occur. The forbidden set can also be interpreted as the subset of
	sequences with infinite positive energy. By applying the map
	$\mathcal{B}_\beta$ to the process and changing $	\beta$ continuously
	from $-\infty$ to $+\infty$ (excluding $\beta = 0$) one can generate any
	rare class of interest $\Lambda^\mathcal{P}_U$. $\beta \to -\infty$
	corresponds to the most probable sequences with the largest energy density
	$U_\text{max}$, $\beta = 1$ corresponds to the typical set and $\beta \to
	+\infty$ corresponds to the least probable sequences with the smallest
	energy density $U_\text{min}$.
	}
\label{fig:bubbles}
\end{figure}

Mirroring the familiar \emph{Boltzmann weight} in statistical physics
\cite{BOLT12}, in the $n \to \infty$ limit, we define the subsets
$\Lambda^\mathcal{P}_U \subset \MeasAlphabet^\infty$ for a process
$\mathcal{P}$ as:
\begin{align}\label{eq:Udef}
&\Lambda^\mathcal{P}_{U,n} && = \left\{w: -\frac{\log_2\mathbb{P}(w)}{n} = U ,
	\ w \in \MeasAlphabet^n \right\}
	 \\
&\nonumber\Lambda^\mathcal{P}_U &&= \lim_{n \to \infty} \Lambda^\mathcal{P}_{U,n}
  ~.
\end{align}
This partitions $\MeasAlphabet^\infty$ into disjoint subsets $\Lambda^\mathcal{P}_U$ in which all $w \in \Lambda^\mathcal{P}_U$ have the same probability decay rate $U$. Physics vernacular would speak of the sequences having the same \emph{energy density} $U$.\footnote{$U$, considered as a random variable, is sometimes called a \emph{self process} \cite{Touc09a}.} Figure~\ref{fig:bubbles} depicts these subsets as ``bubbles'' of equal energy.
Equation~(\ref{eq:TSDEF}) says the typical set is that bubble with energy equal
to the process' Shannon entropy rate: $U = h_\mu$. All the other bubbles
contain rare events, some rarer than others. They exhibit faster or slower
probability decay rates.

Employing a process' HMM to generate realizations produces sequences in the
typical set with probability close to one and, rarely, atypical sequences.
Imagine that one is interested in a particular class of rare sequences, say,
those with energy $U$ ($\Lambda^\mathcal{P}_U$). (One might be concerned about
the class of large-magnitude earthquakes or the emergence of major
instabilities in the financial markets, for example.) How can one efficiently
generate these rare sequences? We now show that there is a new process
$\mathcal{P}^U$ whose typical set is $\Lambda^\mathcal{P}_U$ and this returns
us directly to the challenge of biased sampling.

\section{Biased Sampling}\label{Sec:BiasedSampling}

Consider a finite set of configurations $\{c_i\}$ with probabilities specified
by distribution $\mathbb{P}(\cdot)$ and an associated set $\{\omega_i\}$ of
\emph{weighting factors}. Consider the procedure of reweighting that introduces
a new distribution $\widetilde{\mathbb{P}}(\cdot)$ over configurations where:
\begin{align*}
\widetilde{\mathbb{P}}(c_i) = \frac{\mathbb{P}(c_i)
\exp(\omega_i)}{\sum\limits_i\mathbb{P}(c_i) \exp(\omega_i)}~.
\end{align*}

Given a process $\mathcal{P}$ and its \eM\ $M(\mathcal{P})$, How do we
construct an \eM\ $M(\mathcal{P}^U)$ that generates $\mathcal{P}$'s atypical
sequences at some energy density $U \neq \hmu$ or, as we denoted it, the set
$\Lambda^\mathcal{P}_U$? Here, we answer this question by constructing a map
$\mathcal{B}_\beta: \mathcal{P} \rightarrow \mathcal{P}_\beta$ from the
original process $\mathcal{P}$ to a new one $\mathcal{P}_\beta$. The map is
parametrized by $\beta \in \mathbb{R} /\{0\}$ which indexes the rare set of
interest. (We use $\beta$ for convenience here, but it is related to $U$ by a
function introduced shortly.) Both processes $\mathcal{P}
=\big\{\MeasAlphabet^\infty, \Sigma,\mathbb{P}(\cdot) \big\}$ and
$\mathcal{P}_\beta=\big\{\MeasAlphabet^\infty, \Sigma,\mathbb{P}_\beta(\cdot)
\big\}$ are defined on the same measurable sequence space. The measures differ,
but their supports (allowed sequences) are the same. For simplicity we refer to
$\mathcal{B}_\beta$ as the \emph{$\beta$-map}.

Assume we are given $M(\mathcal{P}) = \big\{ \CausalStateSet, \MeasAlphabet, \{ T^{(\meassymbol)}, \meassymbol \in \MeasAlphabet \} \big\}$. We showed that for every probability decay rate or energy density $U$, there exists a particular $\beta$ such that $M(\mathcal{P}_\beta)$ typically generates the words in $\Lambda^\mathcal{P}_{U,n}$ for large $n$ \cite{Agha17a}. The $\beta$-map which
establishes this is calculated by a construction that relates $M(\mathcal{P})$
to $M(\mathcal{P}_\beta) =\big\{ \CausalStateSet, \MeasAlphabet, \{
\mathbf{S}_\beta^{(\meassymbol)}, \meassymbol \in \MeasAlphabet \}
\big\}$---the HMM that generates $\mathcal{P}_\beta$:
\begin{enumerate}
\setlength{\topsep}{-5pt}
\setlength{\itemsep}{-5pt}
\setlength{\parsep}{-5pt}
\item For each $\meassymbol \in \MeasAlphabet$, construct a new matrix
	$\mathbf{T}^{(\meassymbol)}_\beta$ for which $\big( \mathbf{T}^{(\meassymbol)}_\beta
	\big)_{ij} = \big( \mathbf{T}^{(\meassymbol)} \big)_{ij}^\beta$.
\item Form the matrix $\mathbf{T}_\beta = \sum_{\meassymbol \in \MeasAlphabet}
	T^{(\meassymbol)}_{\beta}$.
\item Calculate $\mathbf{T}_{\beta}$'s maximum eigenvalue $\MaxEigBeta$ and
	corresponding right eigenvector $\MaxRvecBeta$.
\item For each $\meassymbol \in \MeasAlphabet$, construct new matrices
	$\mathbf{S}_\beta^{(\meassymbol)}$ for which:
  \begin{align*}
	 \big( {\textbf S}^{(\meassymbol)}_\beta \big)_{ij}
  	= \frac {\big(\mathbf{T}^{(\meassymbol)}_\beta \big)_{ij} (\MaxRvecBeta)_j }
  	{\MaxEigBeta (\MaxRvecBeta)_i }
  	~.
  \end{align*}
\end{enumerate}

Having constructed the new process $\mathcal{P}_\beta$ by introducing its
generator, we use the latter to produce some rare set of interest
$\Lambda^\mathcal{P}_{U,n}$.

{\The \label{THEO}
In the limit $n \to \infty$, within the new process $\mathcal{P}_\beta$ the
probability of generating realizations from the set
$\Lambda^\mathcal{P}_{U,n}$ converges to one:
\begin{align*}
\lim_{n \to \infty}\mathbb{P}_\beta (\Lambda^\mathcal{P}_{U,n})=1
  ~,
\end{align*}
where:
\begin{align}
U = \beta^{-1} \big(\hmu(\mathcal{P}_\beta) - \log_2 \MaxEigBeta \big)
  ~.
\label{eq:EnergyDensity}
\end{align}
In addition, in the same limit the process $\mathcal{P}_\beta$ assigns equal
energy densities over all the members of the set $\Lambda^\mathcal{P}_{U,n}$.
}

{\ProThe See Ref. \cite{Agha17a}.}

As a result, for large $n$ the process $\mathcal{P}_\beta$ typically generates
the set $\Lambda^\mathcal{P}_{U,n}$ with the specified energy $U$. The process
$\mathcal{P}_\beta$ is sometimes called the \emph{auxiliary}, \emph{driven}, or
\emph{effective} process \cite{Jack10,Garr10a,Chet15}. Examining the form of
the energy, one sees that there is a one-to-one relationship between $\beta$
and $U$. And so, we can equivalently denote the process $\mathcal{P}_\beta$ by
$\mathcal{P}^U$. More formally, every word in $\Lambda^\mathcal{P}_{U}$ with
probability measure one is in the typical set of process $\mathcal{P}_\beta$.  

The $\beta$-map construction guarantees that the HMMs $M(\mathcal{P})$ and
$M(\mathcal{P}_\beta)$ have the same states and transition topology: $\big(
\mathbf{T}^{(\meassymbol)}_\beta \big)_{ij} \neq 0 \iff \big( {\textbf
S}^{(\meassymbol)}_\beta \big)_{ij} \neq 0$. The only difference is in their
transition probabilities. $M(\mathcal{P}_\beta)$ is not necessarily an
\eM---the most memory-efficient classical algorithm that generates the process.
Typically, though, $M(\mathcal{P}_\beta)$ is an \eM\ and there are only
finitely many $\beta$s for which it is not. (More detailed development along
these lines will appear in a sequel.)

\section{Quantum and Classical Costs of Biased Sampling}

Having introduced the necessary background to compare classical versus quantum
models and to appreciate typical versus rare realizations, we are ready to
investigate the quantum advantage when generating a given process' rare events.

The last section concluded that the memory required by the classical algorithm
to generate rare sequences with energy density $U$ is:
\begin{align*}
\Cmu(\Process_{\beta}) = \Cmu(\mathcal{B}_\beta(\Process))
  ~,
\end{align*}
where $U$ and $\beta$ are related via $U = \beta^{-1}
\big(\hmu(\mathcal{P}_\beta) - \log_2 \MaxEigBeta \big)$. Similarly, the memory
required by the quantum algorithm to generate the rare class with energy
density $U$ is:
\begin{align*}
C_q(\mathcal{B}_\beta(\Process))
  ~.
\end{align*}
For simplicity, we denote these two quantities by $\Cmu(\beta) \equiv
\Cmu(\Process_{\beta})$ and $C_q(\beta) \equiv C_q(\Process_{\beta})$.

\begin{figure}
  \centering
\includegraphics[width=0.8\columnwidth]{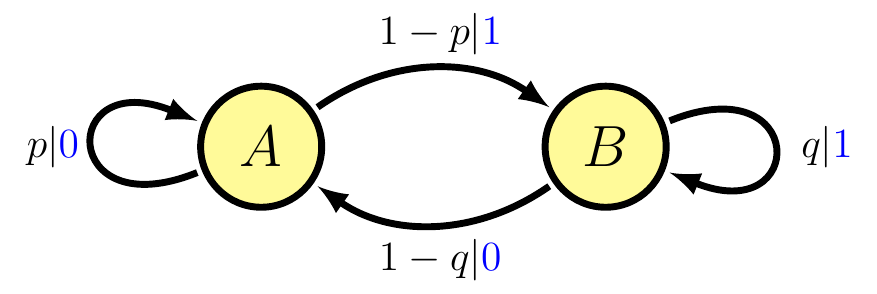}
\caption{\EM\ generator of the Perturbed Coins Process. Edges are labeled with
	conditional transition probabilities and emitted symbols. For example, for
	the self-loop on state $A$, $p|\textcolor{blue}{0}$ indicates the
	transition is taken with probability $\Pr(\textcolor{blue}{0} | A) = p$ and
	the symbol $\textcolor{blue}{0}$ is emitted.
	}
\label{fig:PCSPEM}
\end{figure}

\begin{figure}
  \centering
\includegraphics[width=0.95\columnwidth]{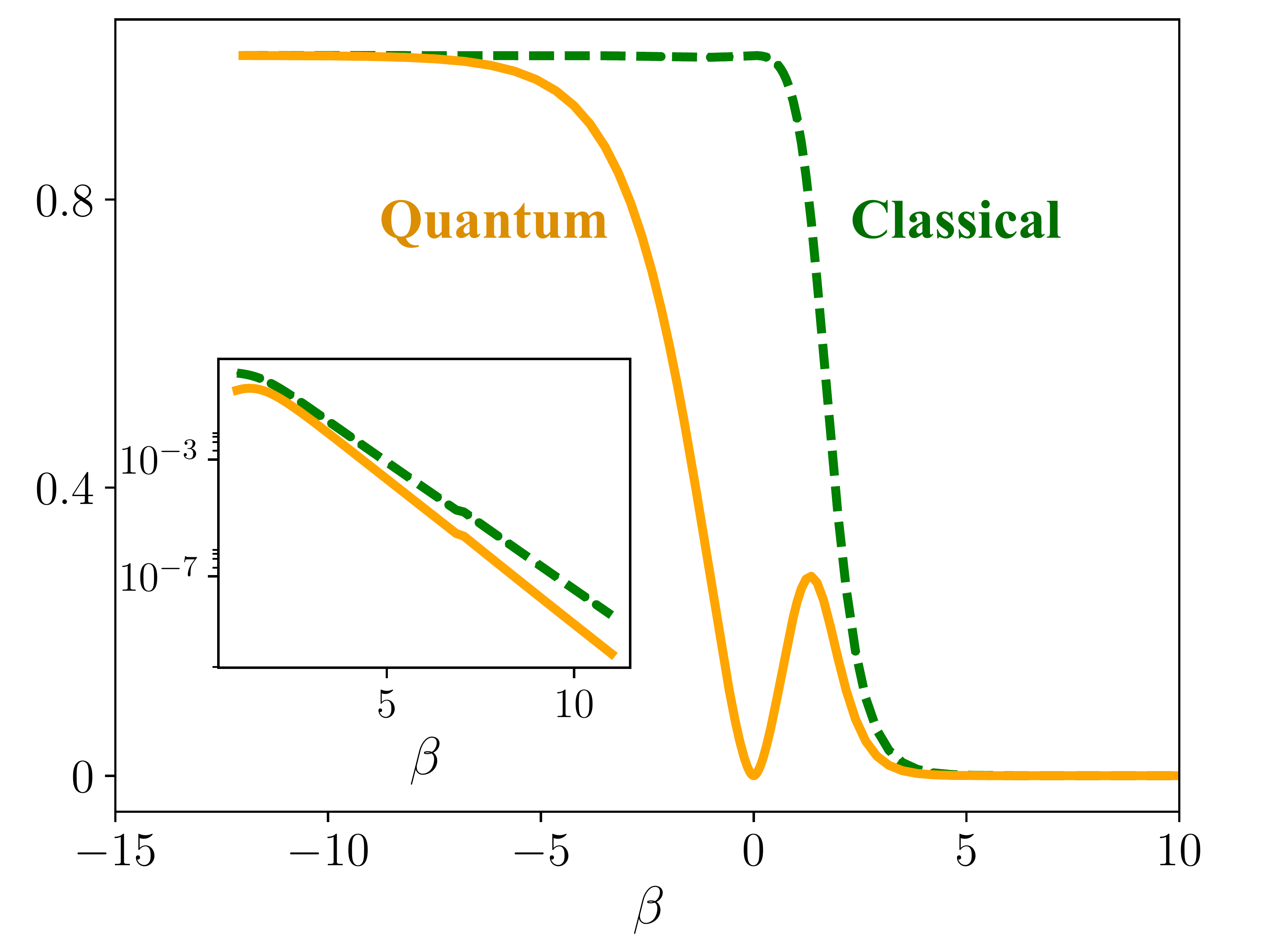}
\caption{Classical memory $\Cmu(\beta)$ and quantum memory $C_q(\beta)$ versus
	$\beta$ for biased sampling of Perturbed Coins Process' rare sequence
	classes: See Fig. \ref{fig:PCSPEM}, with $p=0.6$ and $q=0.8$. As the inset
	shows, for large $\beta$ both classical and quantum memories decay
	exponentially with $\beta$, but the quantum memory decays faster.
	}
\label{fig:CmuCqBeta}
\end{figure}

\subsection{Advantage for a Simple Markov Process}

Consider the case where we have two biased coins, call them $A$ and $B$, and
each has a different bias $p$ and $1-q$ both for Heads (symbol
$\textcolor{blue}{0}$), respectively. When we flip a coin, if the result is
Heads, then on the next flip we choose coin $A$. If the result is Tails, we
choose coin $B$. Flipping the coins over and over again results in a process
$\mathcal{P}^{\text{pc}}$ called the \emph{Perturbed Coins Process}
\cite{Gu12a}. Figure~\ref{fig:PCSPEM} shows the process' \eM\ generator
$M(\mathcal{P}^{\text{pc}})$, where $\CausalStateSet = \{A,B\}$ and
$\MeasAlphabet=\{0,1\}$.

One can also produce this process with a quantum generator
$QM(\mathcal{P}^{\text{pc}})$. Using the construction introduced in
Sec.~\ref{Sec:quantumconstruction}, it has Kraus operators:
\begin{widetext}
\begin{align*}
K_0 \!=\! \frac{1}{d} 
\begin{bmatrix}
\sqrt{q(1-q)p} - p\sqrt{1-p} && p\sqrt{p}-(1-q)\sqrt{p}\\
\sqrt{q(1-q)(1-p)} -(1-p)\sqrt(p) && p\sqrt{1-p}-(1-q)\sqrt{1-p}
\end{bmatrix}
\end{align*}
and:
\begin{align*}
K_1 \!=\! \frac{1}{d} 
\begin{bmatrix}
q\sqrt{(1-q)} - (1-p)\sqrt{1-q} && \sqrt{p(1-p)(1-q)}-(1-q)\sqrt{q}\\
q\sqrt{(q)} -(1-p)\sqrt(q) && \sqrt{p(1-p)q}-q\sqrt{1-q}
\end{bmatrix}
  ~,
\end{align*}
\end{widetext}
where $d=\sqrt{pq} + \sqrt{(1-p)(1-q)}$. For its stationary state distribution
we have:
\begin{align*}
\rho_ s= \frac{1}{2-p-q}
  \begin{bmatrix}
	& 1-p & \alpha \\
	& \alpha & 1-q
  \end{bmatrix}
  ~,
\end{align*}
where $\alpha = (1-q)\sqrt{p(1-p)}+(1-p)\sqrt{q(1-q)}$.

Figure~\ref{fig:CmuCqBeta} shows the classical and quantum memory costs to
generate rare realizations: $\Cmu(\beta)$ and $C_q(\beta)$ versus $\beta$ for
different $\beta$-classes. Surprisingly, the two costs exhibit completely
different behaviors. For example, $\lim \limits_{\beta \to 0}C_q = 0$, while
$\lim \limits_{\beta \to 0} \Cmu = 1$. More interestingly, as the inset
demonstrates, even though both $\Cmu(\beta)$ and $C_q(\beta)$ vanish
exponentially fast, in the limit of $\beta \to \infty$ $C_q(\beta)$ goes to
zero noticeably faster.

We define the \emph{quantum advantage} of biased sampling as the ratio of
classical to quantum memory:
\begin{align*}
\eta(\beta) \equiv \frac{\Cmu(\beta)}{C_q(\beta)}
  ~.
\end{align*}

\begin{figure}
\includegraphics[width=\columnwidth]{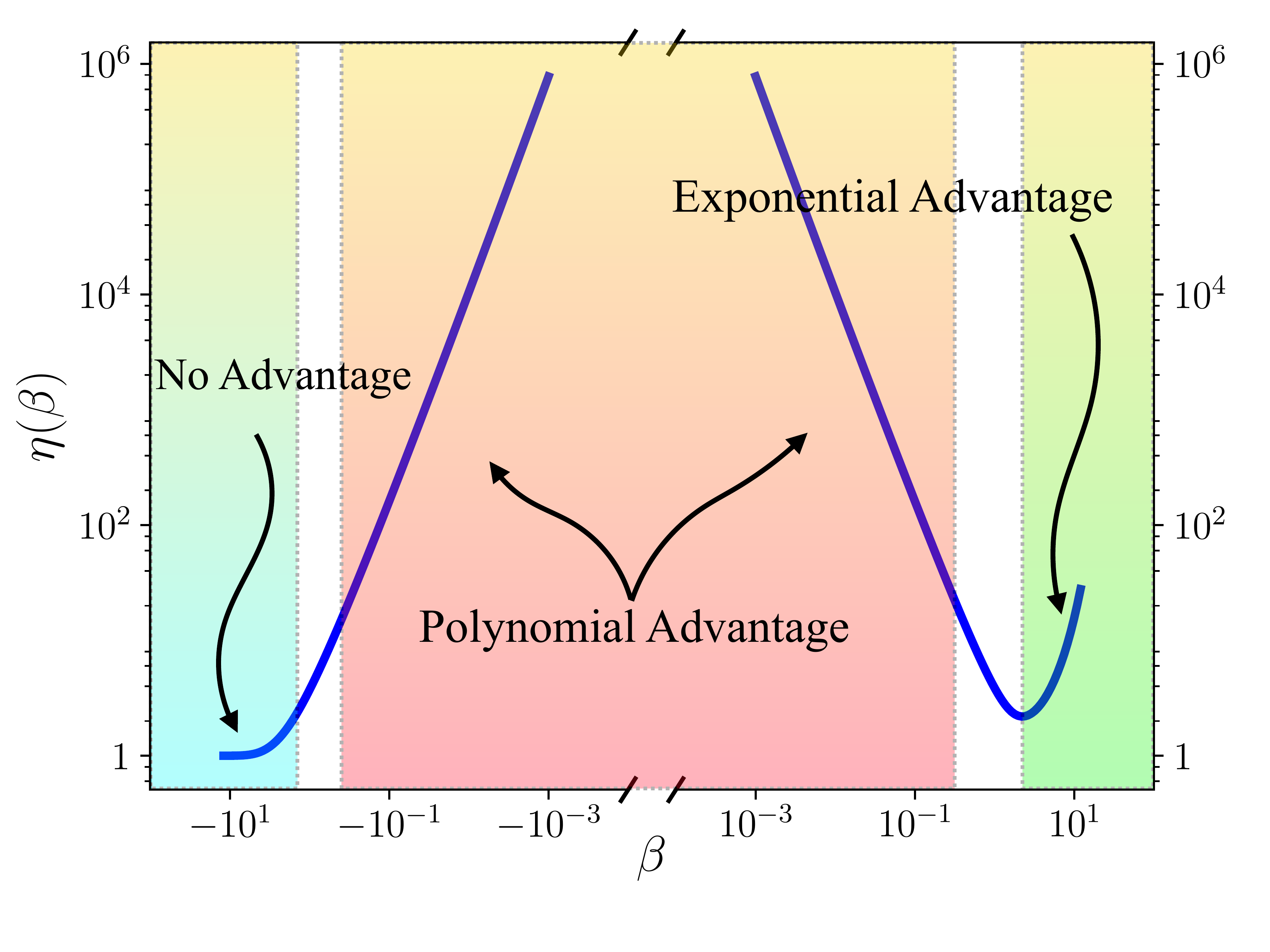}
\caption{Quantum memory advantage for generating the rare realizations of the
	Perturbed Coins Process with $p=0.6$ and $q=0.8$ when employing its
	q-machine instead of it's (classical) \eM. Three different advantages
	occur: (i) near $\beta = 0$ the polynomial advantage scales as
	$\mathcal{O}(\beta^{-2})$, (ii) large positive $\beta$, there is an
	exponential advantage $\mathcal{O}(\exp{(f(q,p)\beta}))$, and (iii) no
	advantage at large negative $\beta$.
	}
\label{fig:EtaBeta}
\end{figure}

Figure~\ref{fig:EtaBeta} graphs the quantum advantage and shows how it divides
into three distinct scaling regimes. First, for small $|\beta|$
(high-temperature) the quantum algorithm exhibits a polynomial advantage
$\mathcal{O}(\beta^{-2})$. Second, for large positive $\beta$ (low-temperature)
the quantum algorithm samples the rare classes with exponential advantage. The
advantage grows as $\mathcal{O}(\exp{(c\beta}))$ as one increases $\beta$ and
where $c$ is a function of $p$ and $q$. Third, for large negative $\beta$
(negative low-temperature regime) there is no quantum advantage. Since we are
analyzing finite-state processes, this regime appears and is the analog of
population inversion. And so, formally there are $\beta$-class events with
negative temperature.

Such is the quantum advantage for the Perturbed Coins Process at $p=0.6$ and
$q=0.8$. The features exhibited---the different scaling regimes---are generic
for any $p>1-q$, though. Moreover, for Perturbed Coins Processes with
$p<1-q$, the positive and negative low temperature behaviors switch.

\subsection{Spin System Quantum Advantage}

Let us analyze the quantum advantage in a more familiar physics setting.
Consider a general one-dimensional ferromagnetic next-nearest-neighbor Ising
spin-\textonehalf\ chain \cite{BAXTER07,Agha12} defined by the Hamiltonian: 
\begin{align}
\mathcal{H}= -\sum_{i} \left( s_is_{i+1} + \tfrac{1}{4} s_is_{i+2} \right)~,
\label{eq:Hamiltonian}
\end{align}
in contact with thermal bath at temperature $\kB T=1$. The spin $s_i$ at site
$i$ takes on values $\{ +1,-1\}$.

After thermalizing, a spin configuration at one instant of time may be thought
of as having been generated left-to-right (or equivalently right-to-left). The
probability distribution over these spatial-translation invariant
configurations defines a stationary stochastic process.
Reference~\cite[Eqs.~$(84)-(91)$]{Feld98b} showed that for any finite and
nonzero temperature $T$, this process has Markov order $2$. More to the point,
the \eM\ that generates this process has $4$ causal states and those states are
in one-to-one correspondence with the set of length-$2$ spin configurations.

\begin{figure}
\includegraphics[width=0.6\columnwidth]{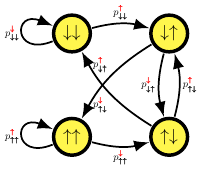}
\caption{\EM\ that generates the spin configurations occurring in the
	one-dimensional ferromagnetic next-nearest-neighbor Ising spin chain with
	the Hamiltonian in Eq.~(\ref{eq:Hamiltonian}).
	}
\label{fig:IsingNNN}
\end{figure}

Figure~\ref{fig:IsingNNN} displays the parametrized \eM\ that generates this
family of spin-configuration processes. To simulate the process, the generator
need only remember the last two spins generated. This means the \eM\ has four
states, $\downarrow\downarrow$, $\downarrow\uparrow$, $\uparrow\downarrow$, and
$\uparrow\uparrow$. If the last two observed spins are $\uparrow\uparrow$ for
example, then the current state is $\uparrow\uparrow$. We denote the
probability of generating a $\color{red}\downarrow$ spin given that the
previous two spins were $\uparrow\uparrow$ by
$p^{{\color{red}\pmb\downarrow}}_{\pmb{\uparrow\uparrow}}$. If the generator is
in the $\uparrow\uparrow$ state and generates a $\color{red}\downarrow$ spin,
then the generator state changes to $\uparrow\downarrow$.

To determine the \eM\ transition probabilities $\{ T^{(\meassymbol)} \}_{
\meassymbol \in \MeasAlphabet}$, we first compute the transfer matrix $V$ for
the Hamiltonian of Eq.~(\ref{eq:Hamiltonian}) at temperature $T$ and then
extract conditional probabilities, following Ref. \cite{Feld98b} and Ref.
\cite{Agha16b}'s appendix. 

\begin{figure}
\includegraphics[width=1\columnwidth]{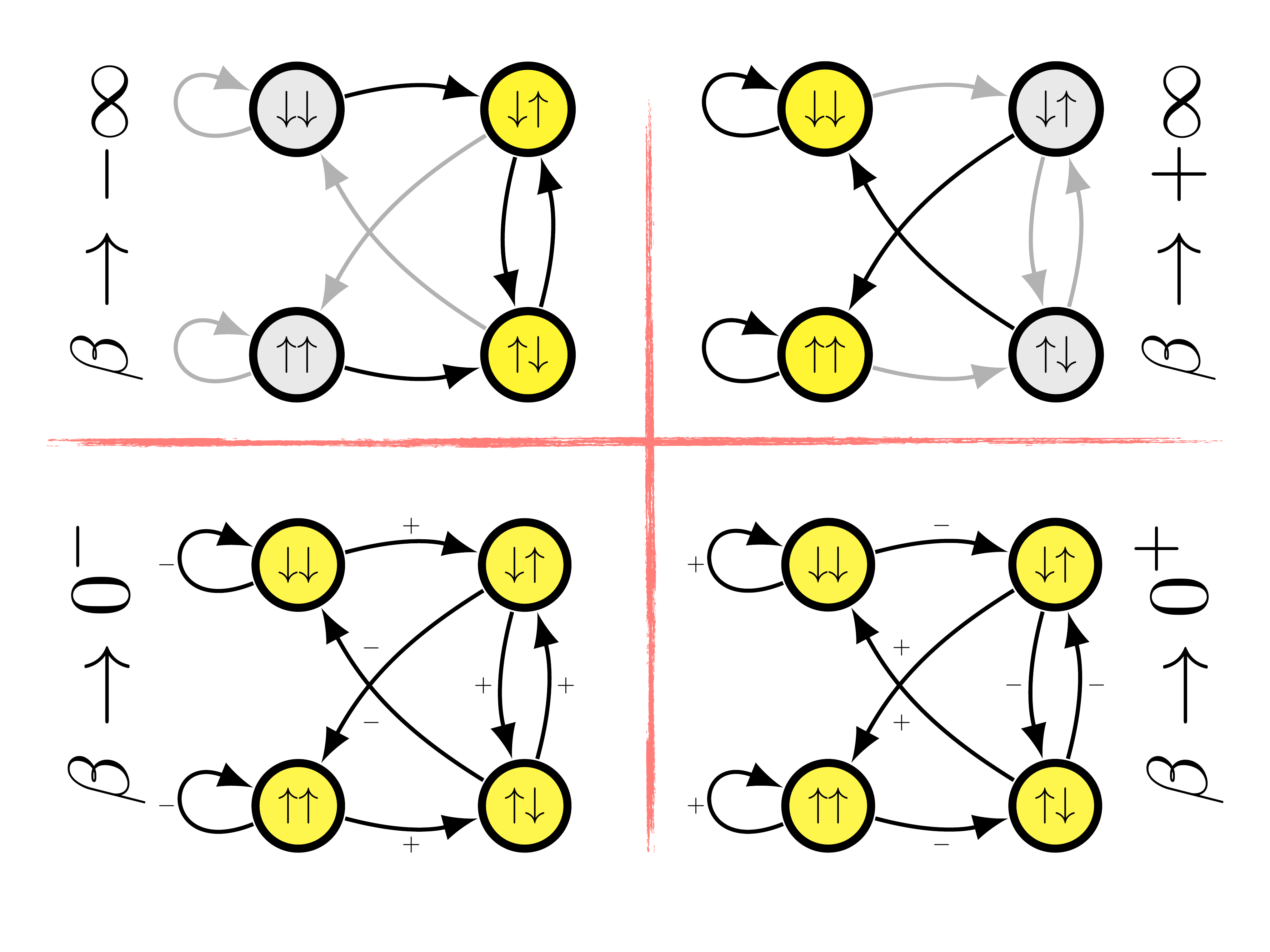}
\caption{Classical generators of four important rare classes:
	(Top-left) Negative zero-temperature limit.
	(Top-right) positive zero temperature limit.
	(Bottom-left) Negative infinite temperature limit.
	(Bottom-right) positive temperature limit.
	Gray edges and states denotes them being rarely visited.
	}
\label{fig:FourLimits}
\end{figure}

\begin{figure}
\includegraphics[width=1\columnwidth]{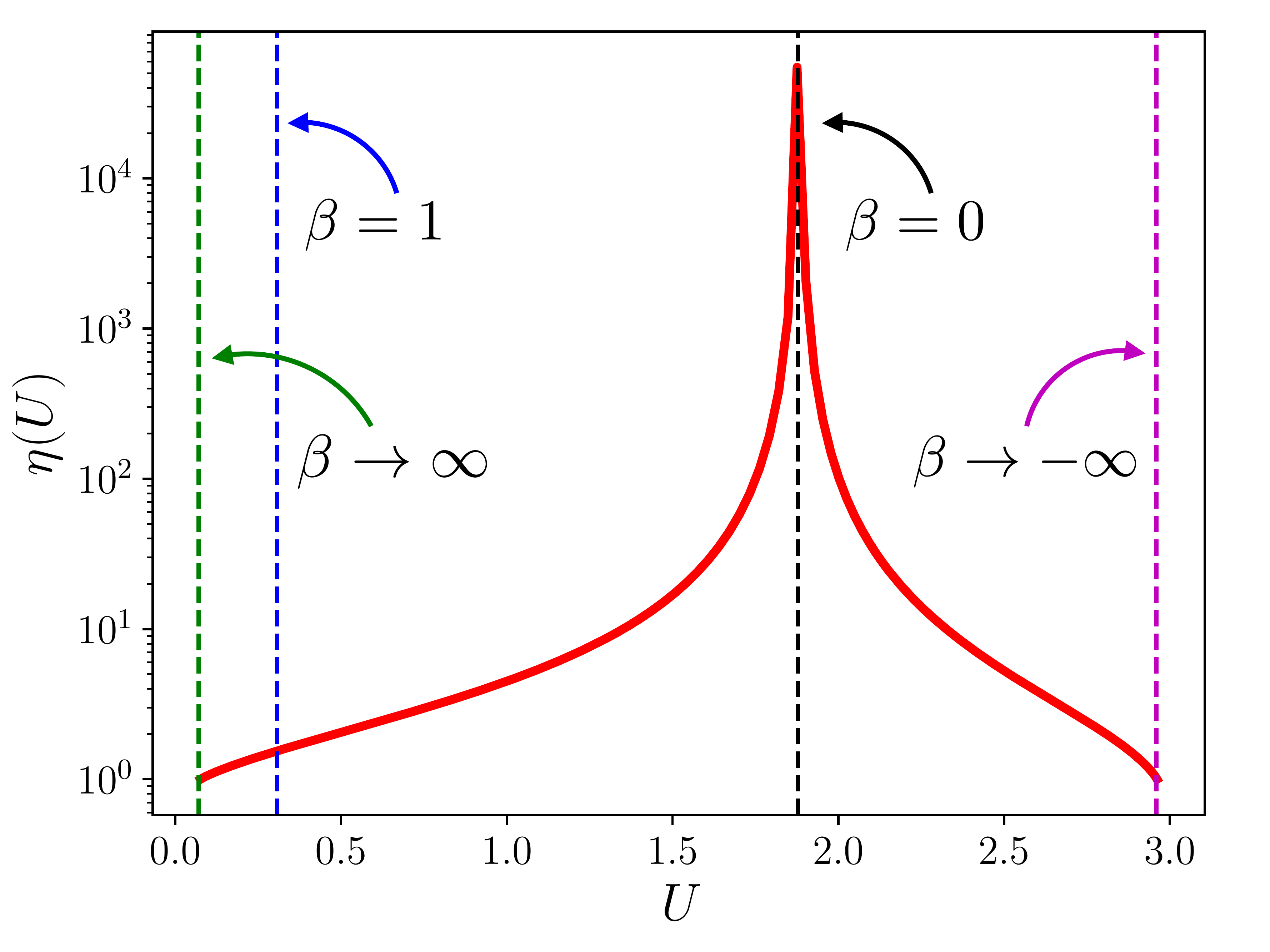}
\caption{Quantum advantage for biased sampling of Ising spin configurations:
	$\eta(U)$ versus decay rate $U$ for bias sampling of equal-energy spin
	configurations. Vertical lines locate $\beta$s corresponding to particular
	$U$s. Note the extreme advantage indicated by the divergence in $\eta(U)$
	at $U = u_0 \approx 1.878$ corresponding to $\beta=0$.
	}
\label{fig:EtaPhase}
\end{figure}

What are the classical and quantum memory costs for bias sampling of the rare
spin-configuration class with decay rate $U$, as defined in
Eq.~(\ref{eq:Udef})? First, note that $U$ is not a configuration's actual
energy density. If we assume the system is in thermal equilibrium and thus
exhibits a Boltzmann distribution over configurations, then $U$ and $E$ are
related via:
\begin{align*}
U = \frac{\log_2(e)}{\kB T} ( E - \mathcal{F}(T))
  ~,
\end{align*}
where:
\begin{align*}
\mathcal{F}(T) = - \kB T \lim_{n \to \infty}\frac{1}{n}
  \ln \left( \sum_{\{w \in \MeasAlphabet^n\}} e^{-\frac{nE(w)}{\kB T}} \right)
  ~.
\end{align*}
This simply tells us that if a stochastic process describes thermalized
configurations of a physical system with some given Hamiltonian, then every
rare-event bubble in Fig.~\ref{fig:bubbles} can be labeled either with $\beta$,
$U$, or $E$. Moreover, there is a one-to-one mapping between every such variable
pair.

Figure~\ref{fig:EtaPhase} plots $\eta(U)$ versus $U$---the quantum advantage of
generating rare configurations with decay rate $U$. To calculate $\eta(U)$ for
a given process $\mathcal{P}$, first we determine the process' classical
generator $M(\mathcal{P})$ using the method introduced in Ref. \cite{Shal98a}.
Second, for every $\beta \in \mathbb{R} /\{0\}$, using the map introduced in
Sec.~\ref{Sec:BiasedSampling}, we find the new classical generator
$M(\mathcal{P}_\beta)$. Third, using the construction introduced in
Sec.~\ref{Sec:QuanAdv}, we find $QM(\mathcal{P}_\beta)$. Fourth, using
Thm.~\ref{THEO} we find the corresponding $U$ for the chosen $\beta$. Using
these results gives $\eta(U) = \Cmu(\beta) / C_q(\beta)$. By varying $\beta$ in
the range $\mathbb{R} /\{0\}$ we cover all the energy density $U$s.
Practically, to calculate $\eta(U)$ in Fig.~(\ref{fig:EtaPhase}), we chose
$2000$ $\beta \in [-10,7.5]$.

As pointed out earlier, $\beta =1$ always corresponds to the process itself.
And, one obtains its typical sequences. As one sees in Fig.~\ref{fig:EtaPhase},
the quantum advantage $\eta(1) < 2$. This simply means that, though there is
a quantum advantage generating typical sequences, it is not that notable.
However, the figure highlights four other interesting regimes.

First, there is the positive zero-temperature limit ($\beta \to \infty$)
corresponding to the rare class with minimum energy density equal to
$U_{\text{min}} = -\log_2
(p^{{\color{red}\pmb\downarrow}}_{\pmb{\downarrow\downarrow}}) = -\log_2
(p^{{\color{red}\pmb\uparrow}}_{\pmb{\uparrow\uparrow}})$. From
Eq.~(\ref{eq:Hamiltonian}) it is easy to see that this rare bubble only has two
configurations as members: all up-spins or all down-spins. Let us consider
finite but large $\beta \gg 1$ that corresponds to the rare class with a low
energy density close to $U_{\text{min}}$. Figure~\ref{fig:FourLimits}(top-left)
shows a general \eM\ for this process. Low color intensity for both edges and
states means that the process rarely visits them during generation. This means,
in turn, that a typical realization consists of large blocks of all up-spins
and all down-spins. These large blocks are joined by small segments.

Second, there is the negative zero-temperature limit ($\beta \to -\infty$) that
corresponds to the rare class with maximum energy density equal to
$U_{\text{max}} = -\frac{1}{2}\log_2
(p^{{\color{red}\pmb\downarrow}}_{\pmb{\downarrow\uparrow}}p^{{\color{red}\pmb\uparrow}}_{\pmb{\uparrow\downarrow}})$.
From Eq.~(\ref{eq:Hamiltonian}) it is easy to see that this rare bubble only
has one configuration as a member: a periodic repetition of spin down and spin
up. Consider finite $\beta \ll 1$ corresponding to a rare class with a high
energy density close to $U_{\text{max}}$.
Figure~\ref{fig:FourLimits}(top-right) shows the general \eM\ for the associated
process. The typical configuration consists of large blocks tiled with spin-up
and spin-down pairs which are connected by other short segments.

Third, there is the positive infinite-temperature limit ($\beta \to 0^+$). In
this limit we expect to see completely random spin-up/spin-down
configurations.  Figure~\ref{fig:FourLimits}(bottom-right) shows the \eM\ for
this class labeled with nonzero small $\beta$. The transition probability for
the edges labeled $+$ is $\half+\epsilon$ and for the edges labeled $-$ is
$\half-\epsilon$, where $\epsilon$ is a small positive number. As one can see,
even though each transition probability is close to one-half, the self-loops
are slightly favored. 

Fourth and finally, there is the negative infinite-temperature limit ($\beta
\to 0^-$). The generator here, Figure~\ref{fig:FourLimits}(bottom-left), is
similar to that at positive infinite temperature, except that the edge-sign
labels are reversed. This means that the self-loops are slightly less favored.

Generating a rare bubble with $\beta<0$ is sometimes called \emph{unphysical
sampling} since there exists no physical temperature at which the system
generates this rare class. As a result, the left part of the
Fig.~\ref{fig:EtaPhase} corresponds to physical sampling and the right part to
unphysical sampling.  That said, there is no impediment to ``unphysical''
sampling from a numerical standpoint. In addition, as we noted, negative
temperatures correspond physically to population inversion, a well-known
phenomenon.

Remarkably, the advantage $\eta(U)$ diverges at $U = u_0 \approx 1.878$,  where
$u_0 = \lim\limits_{\beta \to 0} U$--- both the positive and negative high
temperature limit. Moreover, the advantage $\eta(U)$ diverges as $(U-u_0)^{-2}$
in both limits and, as a result, there is a polynomial-type advantage. For this
specific example one does not find a region with exponential advantage. 

\section{Conclusions}

We introduced a new quantum algorithm for sampling the rare events of classical
stochastic processes. The algorithm often confers a significant memory
advantage when compared to the best known classical algorithm. We explored two
example systems. In the first, a simple Markov process, we found that one gains
either exponential or polynomial advantage. In the second, an Ising chain, we
found a polynomial memory advantage for rare classes in both positive and
negative high-temperature regimes.

Let us address an important point about the optimality of the classical and
quantum algorithms. Consider the integer factorization problem. In this case
Shor's algorithm scales polynomially \cite{Shor99aa}, while the best classical
algorithm currently known scales exponentially \cite{Pome87} with problem size.
While neither algorithm has been proven optimal, many believe that the
separation in scaling is real \cite{Boul16aa}. Similarly, proving optimality
for a rare-event sampling algorithm is challenging in both classical and
quantum settings. However, with minor restrictions, one can show that the
current quantum algorithm is almost always more efficient than the classical
\cite{Maho16}.

\section*{Acknowledgments}
\label{sec:acknowledgments}

The authors thank Leonardo Duenas-Osorio for stimulating discussions on risk
estimation in networked infrastructure. JPC thanks the Santa Fe Institute for
its hospitality during visits as an External Faculty member. This material is
based upon work supported by, or in part by, the John Templeton Foundation
grant 52095, the Foundational Questions Institute grant FQXi-RFP-1609, and the
U. S. Army Research Laboratory and the U. S. Army Research Office under
contracts W911NF-13-1-0390 and W911NF-13-1-0340.

\end{document}